\documentclass[rnote,bibyear]{aa} 

\usepackage{graphicx}
\usepackage{txfonts}
\usepackage{hyperref}
\def\h2{\hskip-2pt}
\begin{document}
   \titlerunning{Average Large separation}
   \title{On the use of the average large separation in  surface layer independent model fitting }
   \author{Ian W. Roxburgh}
   \institute{Astronomy Unit, Queen Mary University of London, 
     Mile End Road, London E1 4NS, UK.
   \email {I.W.Roxburgh@qmul.ac.uk} }
   \date{Received  / Accepted  }
  \abstract{The physics of the outer layers of a star are not well understood but these layers make a major contribution to the  large separation.  We quantify this using stellar models and show that the contribution  ranges from  6\% from the outer 0.1\% of the radius to 30\% from the outer 5\%. and therefore argue that the large separation should not be used as a constraint on surface layer independent model fitting. The mass and luminosity are independent of the outer layers and can be used as constraints, the mass being determined from binarity or from surface gravity and radius.  
 The radius can be used as a constraint but with enhanced error estimates. We briefly consider the determination of the large separation  for  $\alpha$ Cen A and  find that mass derived from surface gravity is  closer to the binary mass than that derived from the large separation.}
      \maketitle

\section{Introduction}
it has long been appreciated that modelling the outer layers of a star is subject to many uncertainties  due to our poor understanding of the physical processes in these layers (cf Christensen-Dalsgaard et al, 1988, Dziembowski et al 1988).  These include
 modelling convection, convective 
overshooting, non-adiabatic effects on both convection and oscillations, turbulent pressure, the equation of state, diffusion, mild turbulence, magnetic fields, rotation and global circulation. All these factors impact on the oscillation frequencies of a model star, and therefore hinder efforts to find stellar models whose frequencies best fit an observed frequency set. 

One way of seeking to overcome these problems is the ``frequency offset technique" (Kjeldsen et al 2008), in which the difference between the observed solar frequencies and those of a ``best solar model" is  scaled by a single factor (determined by the average frequency and large separation) and applied to other stars when seeking a best fit model.  This assumes that the many differences in the properties of the outer layers of  stars  can all be captured in a single scaling factor; such an assumption remains to be verified. 

An alternative approach is to use techniques which are (almost) independent of the structure of the outer layers: these fit
combinations and/or 
properties of the frequencies that only depend on the structure of the inner layers, and consequently can only give information on the interior of a star and not on the outer layers. Such techniques include the ratio of small to large separations (Roxburgh and Vorontsov 2003a, 2013) 
and  matching of phases in the outer layers  (Roxburgh and Vorontsov 2003b, Roxburgh 2010)

When seeking to derive a best fit model for a given set of observed frequencies it is usual to impose some global constraints on the models such as luminosity, effective temperature, surface gravity, radius, and composition, and in particular the average value of the large separation  $\Delta= <\nu_{n+1,\ell} -\nu_{n,\ell}> $ of an observed frequency set  $\{\nu_{n,\ell}\}$, often estimating the mass of the star from the (approximate)   ``scaling relation" $\Delta\propto (M/R^3)^{1/2}$ so that
$${M \over M_\odot}=  \left( \Delta\over\Delta_\odot \right)^2       \left( R \over R_\odot \right)^3  \eqno(1)$$
\noindent where $\Delta_\odot (\approx135\mu$Hz) is the solar value.

As  we show below, the outer layers of a star make a major contribution to the value of the large separation; since the objective of surface-layer independent model fitting  is  to subtract out the effect of the outer layers of the star  it would be inconsistent to constrain the model fitting by requiring that the model had the observed large separation.

\section {Simple analysis}
In the first order asymptotic approximation the oscillation frequencies  satisfy the equation (Vandakurov 1967, Tassoul 1980)
$$\nu_{n,\ell}=\Delta_T ( n + \ell/2 +\epsilon) ~~{\rm where} ~~\Delta_T={1\over 2T}~~{\rm and}~~T=\int_0^R {dr\over c(r)}  \eqno(2)$$
is the acoustic radius of the star, $c$ the sound speed ($c^2=\Gamma_1 P/ \rho$) and $\epsilon$ is a constant. Since $c$ is smallest in the outer layers  the contribution 
of these layers to $T$ and hence $\Delta$ is significant, so uncertainties in the structure of the outer layers can produce significant uncertainty in $\Delta_T$.

To quantify this we separate $T$ into the contribution $t_f$ from the layers below a fractional radius $x_f=r_f/R$,  and $\tau_f=T-t_f$ from the layers above $x_f$. From the definition of $\Delta_T$ (Eqn 2) it follows that
$$\Delta_T = 2\, (t_f + \tau_f) \, \Delta_T^2 \eqno(3)$$
\noindent so the contribution to $\Delta$ from the layers above $x_f$ is $\delta\Delta= 2 \tau_f \Delta_T^2$. 
 In Table 1 we show the contribution $\delta\Delta$ (in $\mu$Hz) from the layers above $x_f$ for a model main sequence  star of $1.10M_\odot$ with central hydrogen abundance $X_c=0.250$ and $\Delta_T=119.6\mu$Hz

\begin{table} [h]
\setlength{\tabcolsep}{5.5pt}
\tiny
\centerline   {\bf Table 1. Contribution {\boldmath $\delta\Delta$ to $\Delta_T$  versus $x_f$~ ($M=1.10M_\odot$)}}
\vskip 0.1cm
\centering
 \begin{tabular}{l r r r r r r  r r} 
\hline\hline 
\noalign{\smallskip}
$x_f$&    0.0 & 0.50&      0.90&      0.95&      0.97&      0.99&   0.995 & 0.999 \\[0.7ex]
$\delta\Delta$& 119.6&    91.16&      45.4&      33.5&      26.4&      15.7&   10.7 &4.4 \\[0.7ex]
 \hline\\  [-1ex]
\end{tabular}
 \end{table} 
As can be seen from Table 1, just the outer 0.1\% of the star  contributes $4\%$  to $\Delta_T$ and the outer 5\% contributes $30\%$, so errors in modelling the outer layers can produce  a substantial change in $\Delta_T$, much greater than  the error estimates 
on the mean large separation of typical frequency sets obtained by the CoRoT or Kepler missions (and of $\alpha$ Cen A\&B and $\beta$ Hydri), which are typically $0.1-0.4 \mu$Hz
(cf Creevey et al 2013).

\section{Full analysis}
The first order asymptotic relation (Eqn 2) is, in general, a poor approximation and the mean large separation $\Delta$ around typical observed $n$ values 
($n\sim 20$) differs from the value given by the 
acoustic radius $\Delta_T$.  For the $1.10 M_\odot$ stellar model $\Delta \approx 117.0\mu$Hz, whereas $\Delta_T=119.6\mu$Hz.  We therefore use a full non-asymptotic analysis to further study 
the effect of uncertainties in the outer layers, replacing the asymptotic relation (2) by the expression  
$$\nu_{n,\ell}=\Delta \,( n + \ell/2 +\epsilon_{n\ell})   \eqno(4)$$
where the {\it phase shifts} $\epsilon_{n\ell}(\nu_{n,\ell})$ are defined by this relation once an average  value of $\Delta$ has been specified. 

As shown by Roxburgh and Vorontsov (2000, 2003a), Roxburgh (2009a), on matching the solution of the oscillation equations integrated away from the centre with the solution integrated in from the surface
at any intermediate acoustic radius $t$, the eigenfrequencies of a star satisfy the equation
$$2\pi T \nu = \pi [ n + \ell/2 ] + \alpha_\ell(\nu,t)  - \delta_\ell(\nu,t)~~{\rm where }~~  T=\int_0^{R_s} {dr\over c}  \eqno(5)$$
is the total acoustic radius of the star (from the centre $r=0$ to the top of the atmosphere $r=R_s$). This equation is identical to Eqn 4, with $\Delta=1/(2T)$ and $\epsilon_{n\ell}=(\alpha_\ell-\delta_\ell)/\pi$.
A different choice of $\Delta$ just adds a term linear in $\nu$ to the $\epsilon_{n\ell}$.

Here $\delta_\ell(\nu,t), \alpha_\ell(\nu,t)$ are inner and outer {\it phase shifts} defined  by the equations
$$  {2\pi\nu\psi\over d\psi/dt } = \tan[2\pi\nu t -\ell\pi/2+\delta_\ell(\nu,t)]~~~~t\le t_f  \eqno(6a)$$
$$  {2\pi\nu\psi\over d\psi/dt } = -\tan[2\pi\nu\tau - \alpha_\ell(\nu,t)]]~~~~~ t\ge t_f \eqno(6b)$$
where  $\psi=r  p' /(\rho c)^{1/2}$ with $p'(r)$ an Eulerian pressure perturbation, $t=\int_0^r dr/c$ the acoustic radius at $r$, $\tau=T-t$\, the acoustic depth, and $t_f$ any arbitrarily chosen acoustic radius.
 For modes of degree $\ell=0, 1$, where the 4th order system of oscillation equation collapse to second order,
the $\alpha_\ell(\nu)$ at any acoustic radius $t_f$ are determined solely by the structure of the layers above $t_f$, and $\delta_\ell(\nu)$ at any $t_f$ are determined solely  by the structure interior to $t_f$.  This is also
a very good approximation for modes of degree $\ell=2,3$ provided $t_f$ is taken in the outer layers where the density is small.

To demonstrate this we take the model of a main sequence star (ModelA) of mass $1.10 M_\odot$, initial composition $X=0.72, Z=0.02$ evolved to a central hydrogen abundance $X_c=0.25$, whose frequencies 
for $\ell=0,1,2$ are shown Fig 1 in a traditional echelle diagram.  
Fig 2 shows the $\alpha_\ell(\nu)$ and $\delta_\ell(\nu)$ for this model at a fitting radius $t_f$, corresponding to a fractional radius $x_f=r_f/R=0.95$. The fact that to high accuracy all the $\alpha_\ell(\nu)$ lie on the same curve independent of $\ell$ is the basis of surface layer independent model fitting techniques.
\begin{figure}[t]
  \begin{center} 
   \includegraphics[width=8.87cm]{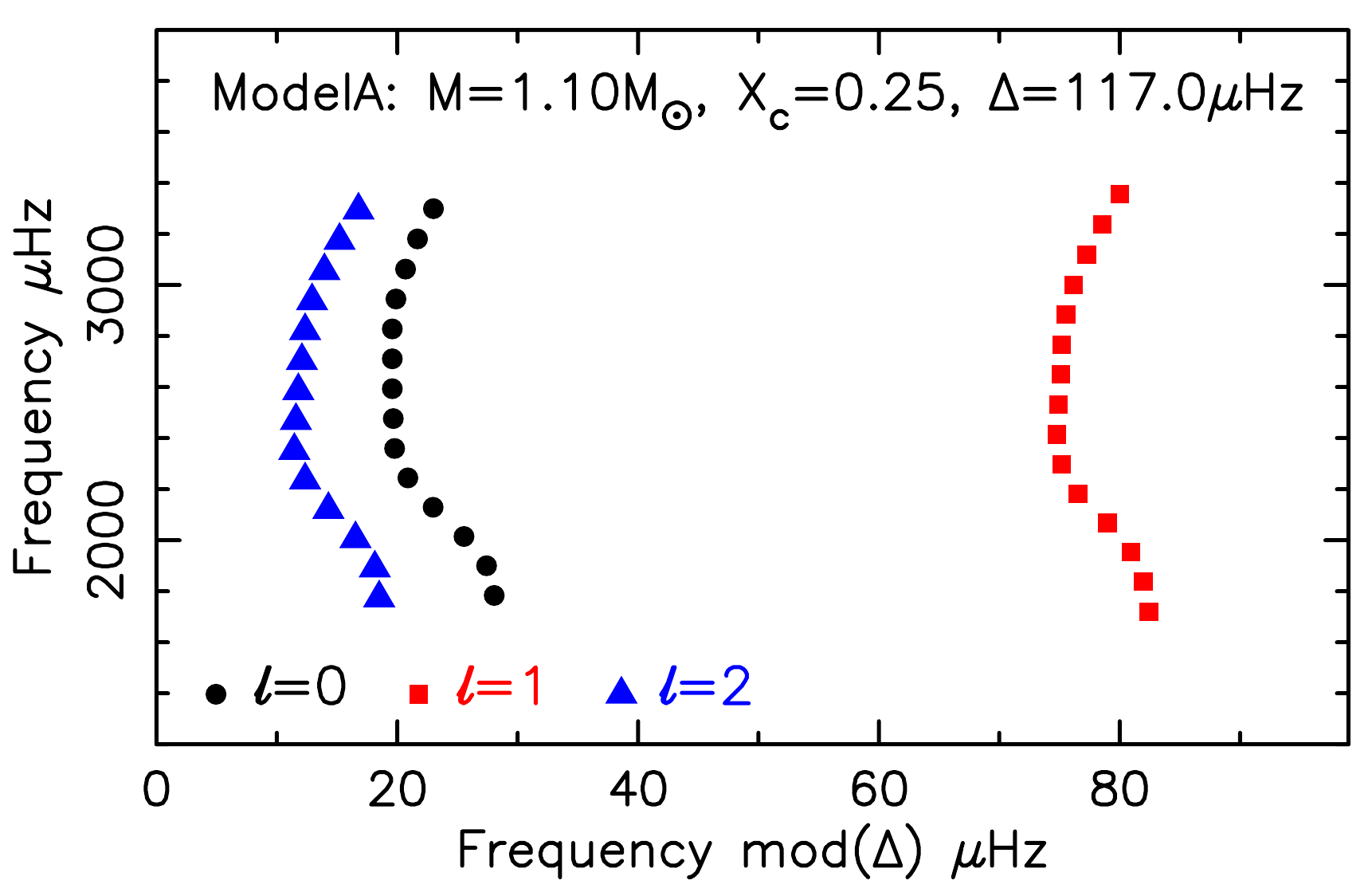}
  \end{center}  
  \vskip-12pt
   \caption{Echelle diagram of the frequencies of modelA, a main sequence star of mass $1.10M_\odot$ evolved to a central hydrogen abundance $X_c=0.25$.}
  \vskip-5pt
  \begin{center} 
   \includegraphics[width=8.87cm]{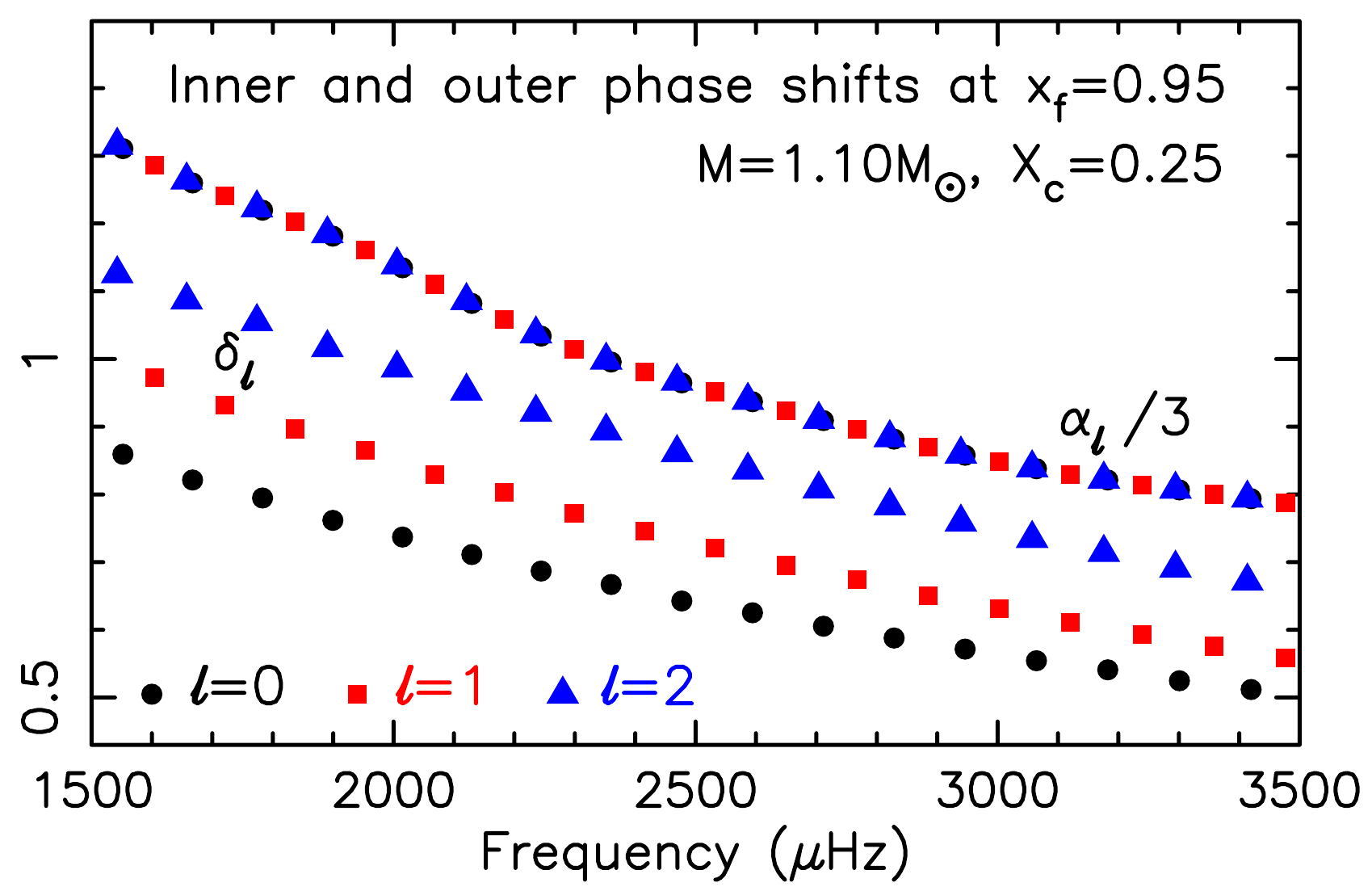}
  \end{center}  
  \vskip-12pt
   \caption{Inner phase shifts $\delta_\ell(\nu)$ and outer phase shifts $\alpha_\ell(\nu)$ for a stellar model of $1.10M_\odot$ and modes
   of degree $\ell=0,1,2$. Note that the $\alpha_\ell(\nu)$ all lie on the same curve.}
  \vskip-12pt
  \end{figure}
  
\
The phases $\alpha_\ell(\nu,t), \delta_\ell(\nu,t)$ can be evaluated for any acoustic radius $t_f$ and any frequency $\nu$, but for an eigenfrequency $\psi$ and $d\psi/dt$ 
must be continuous. Since  $ \tan A + \tan B =0$ gives $\tan(A+B)=0$, it follows that  $A+B=n\pi$, hence, at any $t_f$, an eigenfrequency satisfies the condition
(as in Eqn 5)
$$ 2\pi\nu_{n\ell} \,(t_f  +\tau_f)-\ell\pi/2+\delta_\ell(\nu_{n\ell}, t_f) - \alpha_\ell(\nu_{n\ell},t_f) =n\pi\eqno(7)$$

Subtracting this equation from the corresponding  equation for $n+1$  we obtain
$${1\over \Delta_{n\ell}} =  \left[ \left( 2 t_f +  {1\over\pi}{\partial \delta_\ell \over\partial \nu}\right)_i    +   \left( 2 \tau_f - {1\over\pi} {\partial \alpha_\ell\over\partial \nu}  \right)_o   \right]  \eqno(8)$$
where
$$ { \partial \delta_\ell \over\partial \nu} =  { \delta_\ell(\nu_{n+1,\ell})-\delta_\ell(\nu_{n,\ell} )  \over  \nu_{n+1,\ell}   - \nu_{n,\ell}  }  ~~~~~~
{ \partial \alpha_\ell \over\partial \nu} =  { \alpha_\ell(\nu_{n+1,\ell})-\alpha_\ell(\nu_{n,\ell} )  \over  \nu_{n+1,\ell}   - \nu_{n,\ell}  }  \eqno(9)$$
 The first term in brackets with subscript $i$  in Eqn 8  is determined solely by the structure interior to the fitting point $t_f$ and
 the second term with subscript  $o$ is determined solely by the structure exterior to $t_f$.
On multiplying Eqn 8 by $\Delta_{n\ell}^2$ we deduce that the contribution of the outer layers to $\Delta_{n\ell}$ is 
$$\delta\Delta_{n\ell}(t_f)=    \left( 2\tau_f -  {1\over\pi}{\partial \alpha_\ell \over\partial \nu}\right)  \Delta_{n\ell}^2 \eqno(10)$$

Fig 3 shows $\delta\Delta$  for modes $\ell=0$ and $n=14, 27$, and for different depths,  in terms of the fractional radii $x_f=r/R_p=0.95, 0.97, 0.99, 0.995, 0.999$. 
Here the outer $0.1\%$ of the radius contributes $7\%$ to $\Delta$ and the outer $5\%$ contributes $30\%$, in broad agreement with the results of the 
simple analysis in section 2. So inaccuracies in the modelling of the outer layers can have a large effect on the value of $\Delta$.
\begin{figure}[t]
  \begin{center} 
   \includegraphics[width=8.87cm]{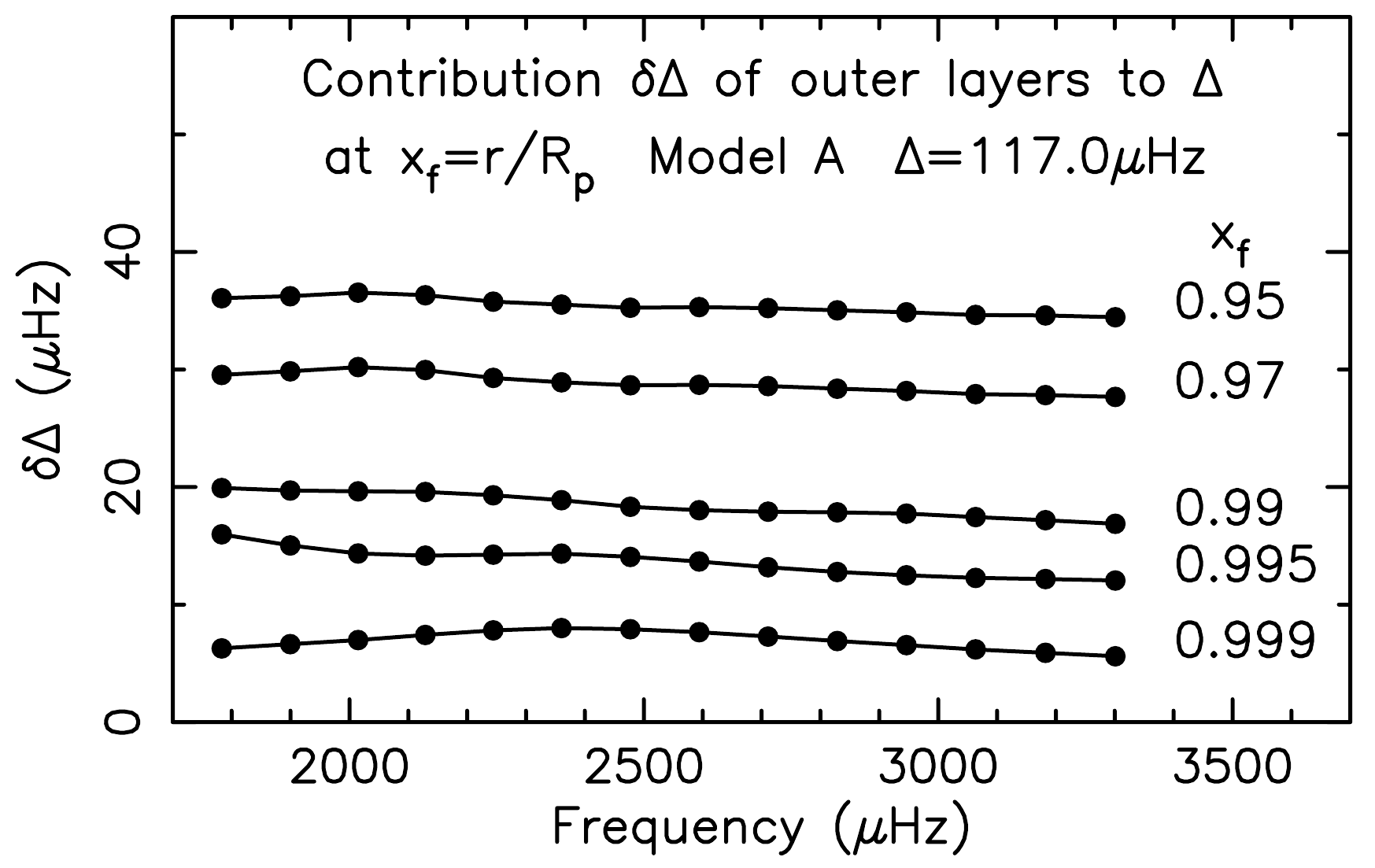}
  \end{center}  
  \vskip-12pt
   \caption{ Contribution $\delta\Delta$ of the outer layers to the large separations $\Delta_{n,0}$ 
    for different depths, for a stellar model of $1.10 M_\odot$ with $\Delta \sim 117\mu$Hz. $R_p$ is the photospheric radius.}
  \vskip-12pt
\end{figure}

These results apply equally to more evolved stars. To show this we undertook the same analysis for model B, a highly evolved post main sequence star  of mass $1.15M_\odot$  with initial composition $X=0.72,\,Z=0.015$  in the shell burning phase moving 
over to the red giant branch.  The model has many mixed modes as can be seen in the echelle diagram (Fig 4). The phase shifts are shown in Fig 5:  the inner phase shifts
 $\delta_{n\ell}$ for $ \ell=1,2$ no longer lie on smooth curves but the outer phase shifts $\alpha_{n\ell}$ still all lie on a single curve for all $\ell$. In Fig 6 we give the contribution of the outer layers to the large separation $\Delta$ for $14~ \ell=0$ modes with $n=9,22$, as a function of fractional radius $x_f$.  Here the outer $0.1\%$ of the radius contributes $6.5\%$ to $\Delta$ and the 
outer $5\%$ some $32\%$, more or less the same as for Model A.

\begin{figure}[t]
  \begin{center} 
   \includegraphics[width=8.87cm]{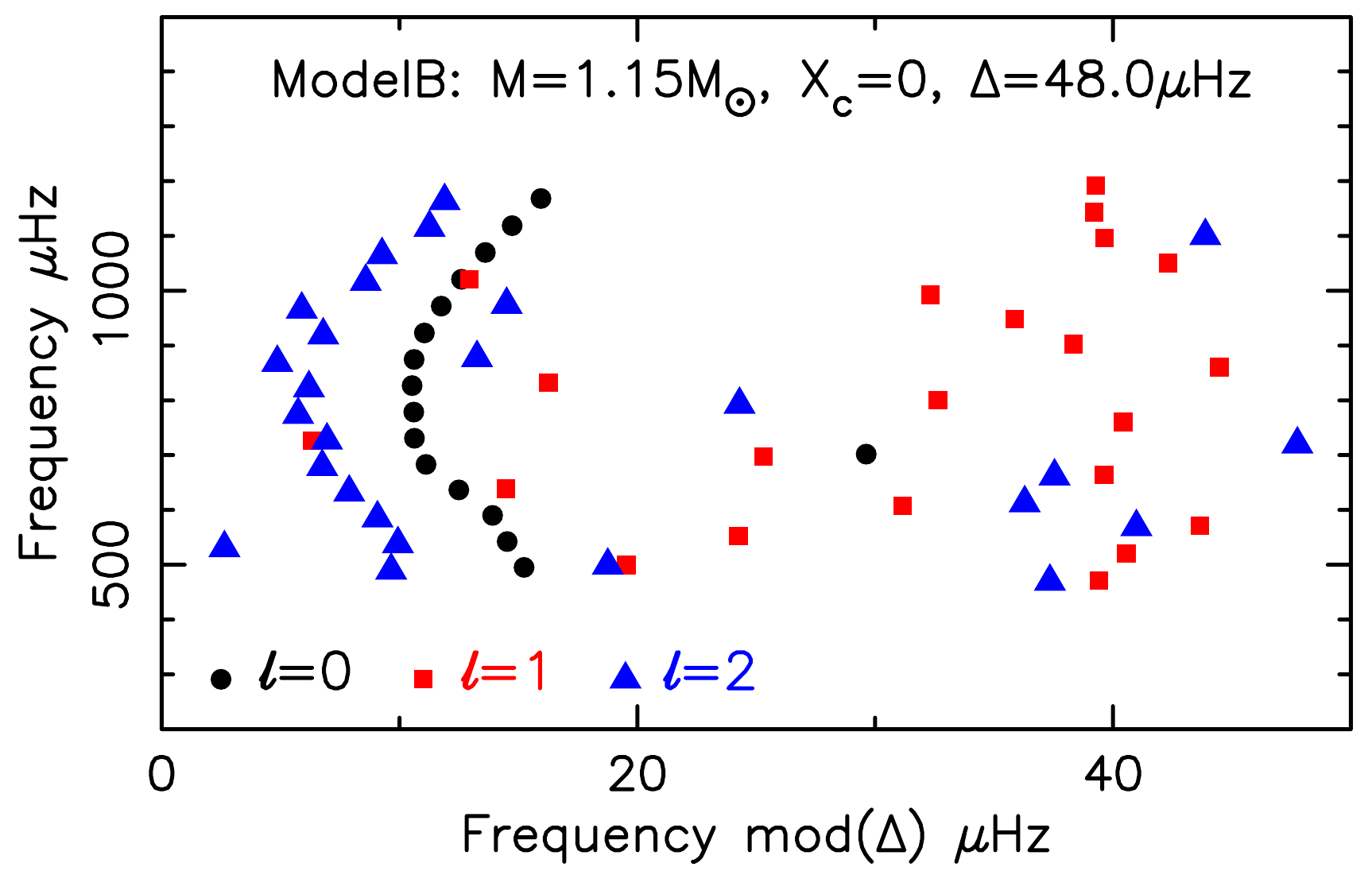}
  \end{center}  
  \vskip-12pt
   \caption{Echelle diagram of the frequencies of modeB, a $1.15M_\odot$ post main sequence model in the  shell burning  phase - the model has many mixed modes for both $\ell=1$ and $\ell=2$. }
  \vskip-5pt
  \begin{center} 
   \includegraphics[width=8.87cm]{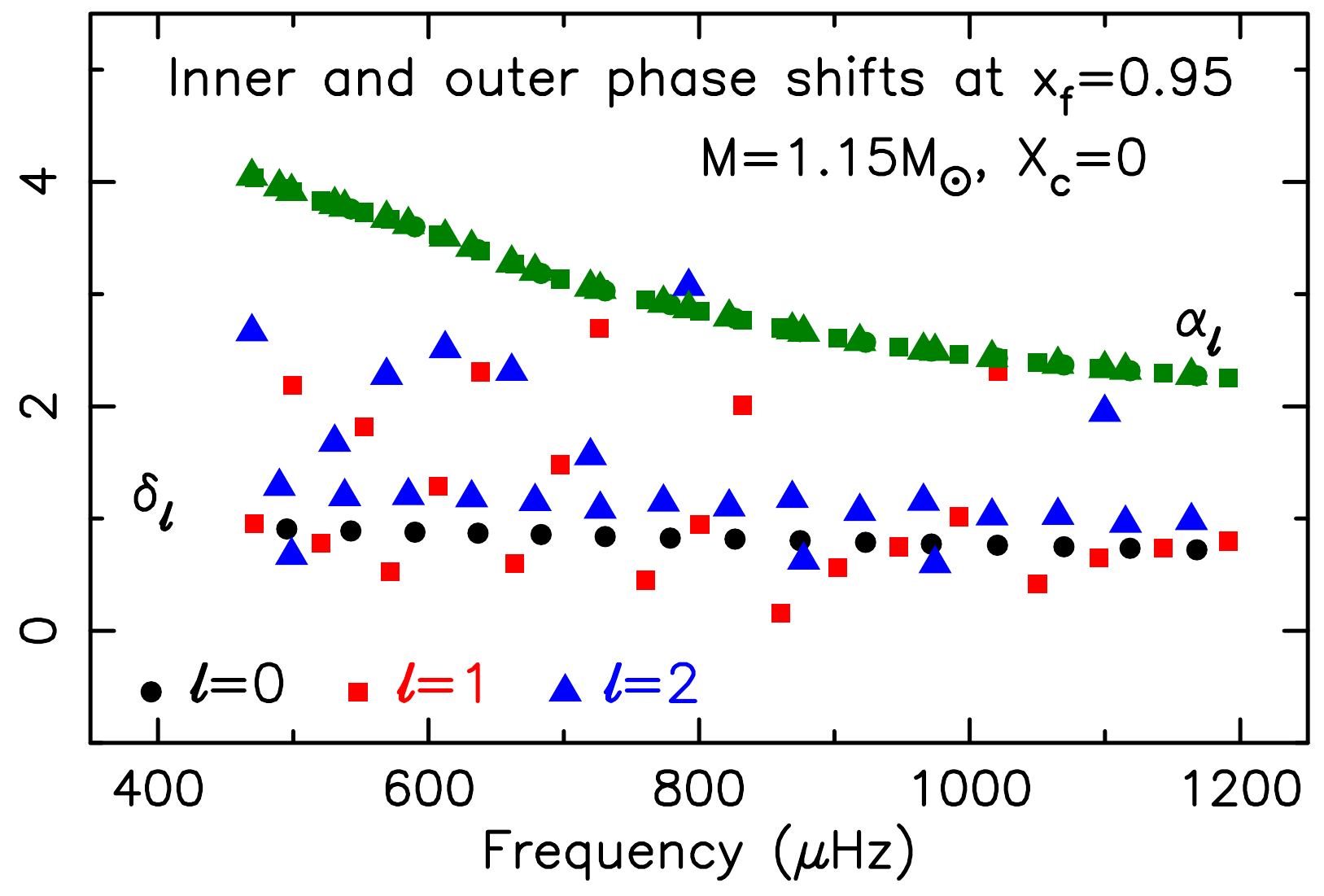}
  \end{center}  
  \vskip-12pt
   \caption{Inner phase shifts $\delta_\ell(\nu)$ and outer phase shifts $\alpha_\ell(\nu)$ for a stellar model of $1.15M_\odot$ and modes
   of degree $\ell=0,1,2$. Note that the $\alpha_\ell(\nu)$ (in green) still all lie on a single curve.}
  \vskip-12pt
\end{figure}
  \begin{figure}[t]
  \begin{center} 
   \includegraphics[width=8.87cm]{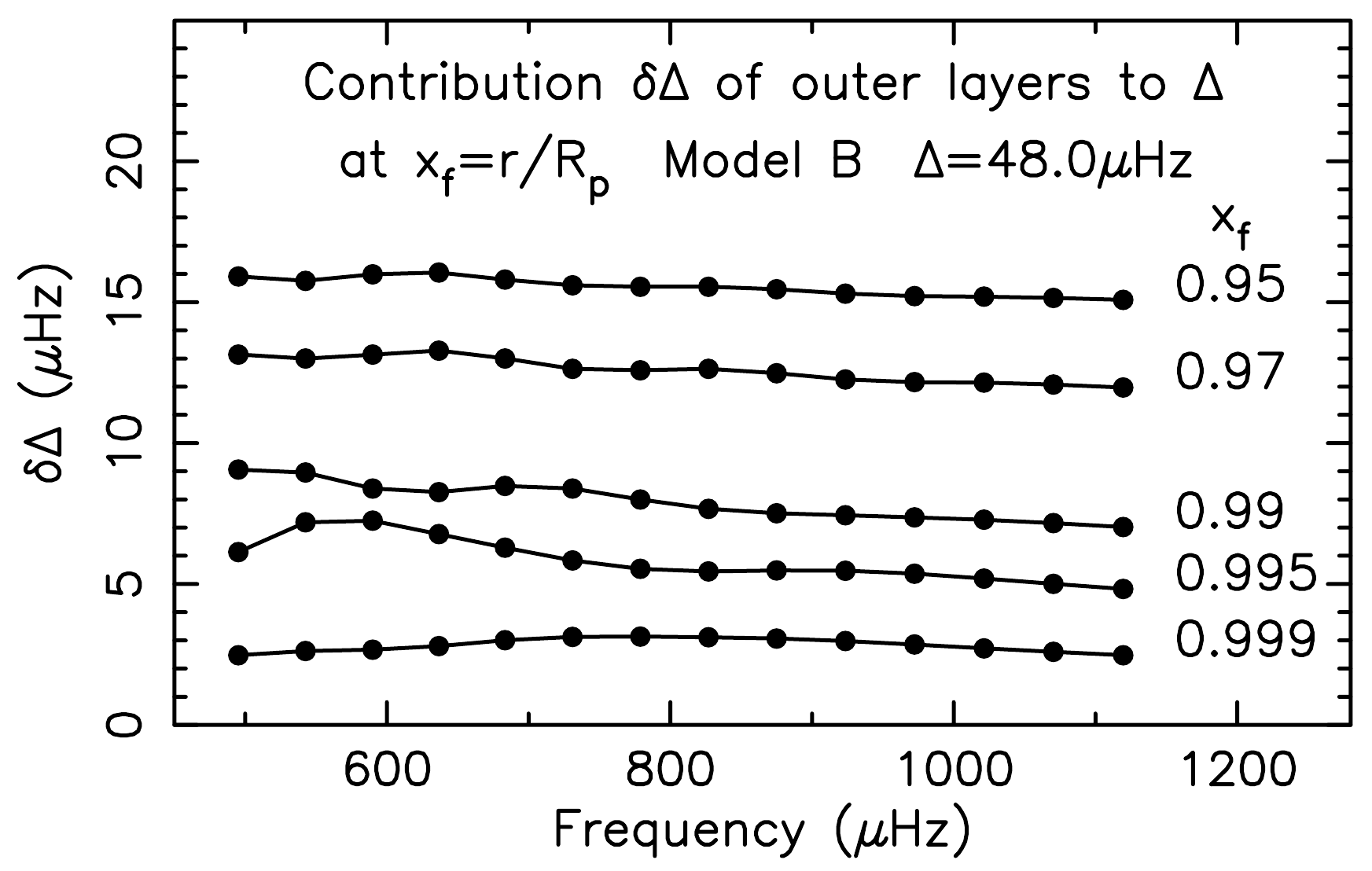}
  \end{center}  
  \vskip-12pt
   \caption{ Contribution $\delta\Delta$ of the outer layers to the large separations $\Delta_{n,0}$ 
    for different depths, for a modelB  of $1.15 M_\odot$ with $\Delta \sim 48\mu$Hz. $R_p$ is the photospheric radius.}
  \vskip-12pt
\end{figure}

\section{Global constraints on surface layer independent model fitting}
The above analysis demonstrates that the even small differences in the structure of the outer layers of a star can make a significant difference to the value of the large separation
$\Delta$.  When using surface layer independent model fitting (separation ratios or phase matching), one is seeking to subtract out the effect of the outer layers so it is inconsistent to 
constrain the search by requiring the model fit the observed large separation. 

One can ask what global constraints should one  impose.  The two obvious global constraints are the mass $M$ and luminosity $L$, since these are essentially determined solely by the inner structure of the star.  If the star has a measured parallax then the luminosity can be estimated, but the mass is only known for stars in binary systems, the prime example being
$\alpha$ Cen A\&B. 

In principle one can estimate the mass from spectroscopically determined surface gravity $g$, and surface radius $R$ 
estimated either from interferometry or from $L, T_{eff}$.  Such a mass estimate  is independent of
uncertainties due to surface layer contributions to $R$, but often $g$ is not known to high enough precision to provide a useful constraint.
The outer layer contributions to $R$ are much smaller that those to $\Delta$; the layers above $x_f=0.99$ contribute $1\%$ to the radius
but $17\%$ to the large separation,  so one could reasonably impose a radius constraint but with an enhanced error estimate to allow for the unknown contribution  of the outer layers. 

\section{{\boldmath $\alpha$ Cen A }}
The fundamental properties of $\alpha$ Cen A have been determined to high precision; these and the derived radius are listed in Table 2: the binary mass is from Pourbaix et al. (2002);  parallax from Soderhjelm (1999); angular diameter from Kervella  et al (2003); $\log\,g$ from Bruntt et al (2010).

\begin{table} [h]
\setlength{\tabcolsep}{2.7pt}
\tiny
\centerline   {\bf Table 2. {\boldmath $\alpha$ Cen A\&B: observational input }}
\vskip 0.11cm
\centering
 \begin{tabular}{l l c c c c c c c c c c c } 
\hline\hline  \\[ -1ex]
 & M/M$_\odot$(B)$$&$\pi\, ({\rm mas})$& $\theta\, ({\rm mas})$ & $\log g$\,(cm/s$^2$)&R/R$_\odot$ \\	[1ex]
\hline 
\noalign{\smallskip}
 A&$1.105 \pm0.007$& $747.1\pm 1.2$& $8.511 \pm 0.020$ & $4.309\pm0.055$& $1.224\pm0.003$\\ [0.1ex]
\hline\\  [-1ex]
\end{tabular}
\smallskip
 \end{table} 
\vskip -6pt
There have been several investigations to determine the frequencies of $\alpha$ Cen A (Bouchy and Carrier 2002, Bedding et al 2004, Fletcher et al, 2006, Bazott al 2007), and more recently by de Meulenaer et al (2010) who combined the time series from Bouchy and Carrier with those from Bedding.  The different investigations are not in total agreement with each other indicating either the difficulty in extracting frequencies and estimating uncertainties, or possibly a variation in time, or both.  The average large separation $\Delta$ estimated from these frequencies  are all in the  vicinity of $106\mu$Hz but vary by $\sim 0.5\mu$Hz.

We here estimate the average large separation directly from the widowed autocorrelation of the de Meulenaer combined time series as this does not depend on uncertainties in frequency determination, and is more robust since it combines all the data into the determination of the one quantity - the average $\Delta$ (Roxburgh and Vorontsov 2006, Roxburgh 2009b).  To be specific we used a $\sin^2$ window of FWHM =8$\Delta$ centred on the peak of maximum power, here found to be $\nu_{max}=2384\mu$Hz. 
 The resulting $\Delta=106.1\mu$Hz.  The value varies with the choice of window width and $\nu_{max}$ by $\sim 0.2\mu$Hz which we take as an error estimate on $\Delta$.  We did the same analysis using the SPM solar time series (kindly supplied by T Appourchaux) to determine the solar value of $\Delta_\odot=134.97 \pm 0.1\mu$Hz.  


\begin{table} [h]
\setlength{\tabcolsep}{3.5pt}
\tiny
\centerline   {\bf Table 3. {\boldmath $\alpha$ Cen A\&B:  derived radii and masses } }
\vskip 0.15cm
\centering
 \begin{tabular}{l l c c c c c c c c c c c } 
\hline\hline\\ [-1ex]
 & M/M$_\odot$(B)& $\Delta(\mu$Hz) & M(g)/M$_\odot$& M($\Delta$)/M$_\odot$ \\	[1ex]
\hline 
\noalign{\smallskip}
 A&$1.105 \pm0.007$&$106.1\pm0.2$&$1.115\pm 0.151$&$1.134 \pm  0.011$ \\ [0.5ex]
\hline\\ [-0.4ex]
\end{tabular}
\smallskip
\vskip-8pt
 \end{table} 
 
 The resulting mass $M(\Delta)$ obtained from the large separation $\Delta$ using the scaling relation (Eqn 1) is shown in Table 3 together with the mass $M(g)$ 
 derived from $\log g$. (We took the following solar values:  M$_\odot$=1.98855\,10$^{33}$gm, R$_\odot$=6.9599\,10$^{10}$cm.) 

It is interesting to note that mass $M(g)$  is more compatible with the dynamically determined value than is  $M(\Delta)$, or equivalently that the spectroscopic $\log g$ is closer to the dynamical value than is the value obtained from the large separation scaling relation (Eqn 1).

 \newpage
Not too much weight should be given to this result as the combined time series is still only $11.4$ days in duration so the accuracy of our value of $\Delta$ is open to question.  To address this issue  we determined the values of the average solar large separation $\Delta_\odot$ from 50 non overlapping $11.4$ day time series  taken from the SPM data.  Here we found that $\Delta_\odot$ varied between  $134.66$ and $135.31\mu$Hz indicating an uncertainty of $\pm0.4\mu$Hz 
whereas the same analysis  on 20 non overlapping 150 day time strings found  a variation between $134.87$ and $135.06$, consistent with the previously estimated 
uncertainty of $0.1\mu$Hz.  This suggests that the uncertainty on the $11.4$ day estimate of 
$\Delta$ for $\alpha$ Cen A  should be enhanced by a factor $\sim 4$ to $0.8\mu$Hz which would give an estimate of 
 $M(\Delta)/M_\odot = 1.134 \pm 0.020$  almost within overlapping $1\sigma$ errors  of the binary mass of  $M_A/M_\odot=1.105\pm0.007$.

This also suggests that the values of the frequencies and their error estimates for $\alpha$ Cen A could be different for different short duration time series,
which may in part explain the difference between the estimated frequencies obtained by different authors.

\section*{Acknowledgements}
The author thanks T Appourchaux for providing the SPM time series data and P de Meulenaer for providing the merged $\alpha$ Cen A time series data.
The author gratefully acknowledges support from the Leverhulme Foundation under grant EM-2012-035/4

\end{document}